\documentclass[fleqn,usenatbib,useAMS]{mnras}

\newcommand{\host}{EPIC~210363145}

\newcommand{\rsun}{$R_{\odot}$}
\newcommand{\rearth}{$R_{\oplus}$}

\newcommand{\kep}{{\it Kepler }}
\newcommand{\logg}{log~$g$}

\newcommand{\um}{$\mu$m}

\newcommand{\mps}{m~sec$^{-1}$}
\newcommand{\msun}{$M_{\odot}$}

\newcommand\kms{km~s$^{-1}$}

\newcommand\teff{\ensuremath{T_\text{eff}}}

\usepackage{longtable}
\usepackage{lscape}
\usepackage{graphicx}	% Including figure files
\usepackage{amsmath}	% Advanced maths commands
\usepackage{amssymb}	% Extra maths symbols
\usepackage{multicol}        % Multi-column entries in tables
\usepackage{bm}		% Bold maths symbols, including upright Greek
\usepackage{pdflscape}	% Landscape pages

\usepackage[T1]{fontenc}
\usepackage{ae,aecompl}

\usepackage{txfonts}

\title[]{Zodiacal Exoplanets in Time (ZEIT) II.  A "Super-Earth"\\Orbiting a Young K Dwarf in the Pleiades Neighborhood}

\author[Gaidos et al.]{E. Gaidos,$^{1,2}$\thanks{E-mail: gaidos@hawaii.edu} A.~W. Mann,$^{3,4}$ 
A. Rizzuto,$^{3}$ L. Nofi,$^{5}$ G. Mace,$^{3}$ A. Vanderburg,$^{6}$ \newauthor G. Feiden,$^{7}$ N. Narita, $^{8,9,10,11}$, Y. Takeda,$^{10}$ T. M. Esposito,$^{12}$ R. J. De Rosa,$^{12}$ \newauthor M. Ansdell,$^{5}$ T. Hirano,$^{13}$ J. R. Graham,$^{12}$ A. Kraus,$^{3}$ D. Jaffe$^{3}$
\\
$^{1}$Department of Geology \& Geophysics, University of Hawaii at M\={a}noa, Honolulu, Hawaii USA\\
$^{2}$Visiting Scientist, Center for Space \& Habitability, University of Bern, Switzerland\\
$^{3}$Department of Astronomy, University of Texas at Austin, Austin, Texas USA\\
$^{4}$NASA Hubble Postdoctoral Fellow\\
$^{5}$Institute for Astronomy, University of Hawaii at M\={a}noa, Honolulu, Hawaii USA\\
$^{6}$Harvard-Smithsonian Center for Astrophysics, Cambridge, Massachusetts USA\\
$^{7}$Department of Astronomy, Uppsala University, Upsalla, Sweden\\
$^{8}$Department of Astronomy, The University of Tokyo, Bunkyo-ku, Tokyo, Japan\\
$^{9}$Astrobiology Center, National Institutes of Natural Sciences, Mitaka, Tokyo, Japan\\
$^{10}$National Astronomical Observatory of Japan, Mitaka, Tokyo, Japan\\
$^{11}$SOKENDAI (The Graduate University for Advanced Studies), Mitaka, Tokyo, Japan\\
$^{12}$Department of Astronomy, University of California at Berkeley, Berkeley, California USA\\
$^{13}$Department of Earth and Planetary Sciences, Tokyo Institute of Technology, Meguro-ku, Tokyo, Japan\\
}

\begin{document}
\date{Submitted to MNRAS 18 June 2016}
\pagerange{\pageref{firstpage}--\pageref{lastpage}} \pubyear{2016}

\maketitle
\label{firstpage}

\begin{abstract}
We describe a "super-Earth"-size ($2.30 \pm 0.16$\rearth{}) planet transiting an early K-type dwarf star in the Campaign 4 field observed by the K2 mission.  The host star, \host{}, was identified as a member of the approximately 120-Myr-old Pleiades cluster based on its kinematics and photometric distance.  It is rotationally variable and exhibits near-ultraviolet emission consistent with a Pleiades age, but its rotational period is $\approx$~20~d and its spectrum contains no H$\alpha$ emission nor the Li I absorption expected of Pleiades K dwarfs.  Instead, the star is probably an interloper that is unaffiliated with the cluster, but younger ($<1$~Gyr) than the typical field dwarf.  We ruled out a false positive transit signal produced by confusion with a background eclipsing binary by adaptive optics imaging and a statistical calculation.  Doppler radial velocity measurements limit the companion mass to $<2$ times that of Jupiter.  Screening of the lightcurves of 1014 potential Pleiades candidate stars uncovered no additional planets.  An injection-and-recovery experiment using the K2 Pleiades lightcurves with simulated planets, assuming a planet population like that in the \kep{} prime field, predicts only 0.8-1.8 detections (vs. $\sim 20$ in an equivalent \kep{} sample).  The absence of Pleiades planet detections can be attributed to the much shorter monitoring time of K2 (80 days vs. 4 years), increased measurement noise due to spacecraft motion, and the intrinsic noisiness of the stars.  
\end{abstract}


%%%%%%%%%%%%%%%%%%%%%%%%%%%%%%%%%%%%%%%%%%%%%%%%%%

%%%%%%%%%%%%%%%%% BODY OF PAPER %%%%%%%%%%%%%%%%%%

\section{Introduction}

Since antiquity, the Pleiades cluster has been used to mark the passage of time.  The name may derive from the ancient Greek word $\pi \lambda \epsilon \omega$ ("to sail") because the constellation's helical rising marked the advent of fair weather sailing in the Mediterranean.\footnote{The helical rising of Pleiades is currently the first week of June but due to the precession of the equinoxes would have occurred mid-spring in Iron Age Europe.}  The appearance of the Pleiades (Makali'i) in the evening sky begins Makahiki, the Hawaiian harvest season.  

Scientific studies of the Pleiades revealed it to be a nearby cluster of very young \citep[e.g., $112\pm5$~Myr][]{Dahm2015} stars and this object serves studies of star formation, stellar evolution, stellar dynamics, ultracool stars and planetary-mass bodies \citep{Casewell2007,Zapatero2014}.  It provides a (controversial) calibration tie-point for distance measurements \citep{Melis2014}.  Observations of the Pleiades, and other young clusters are also a way to identify and study planets younger than those in surveys of the field such as \kep{}.  Many important events probably occur in the first few hundred million years (Myr) of planetary systems, e.g. thermal and radius evolution of gas- and ice-giant planets \citep{Fortney2011,Guillot2014}, the heating and escape of planetary atmospheres driven by elevated ultraviolet (UV) emission from the active host star \citep{Tian2005}, planet migration, orbital resonance crossing \citep{Gomes2005,Thommes2008}, and collisions \citep{Volk2015}.  Comparisons of the planet populations in young clusters such as the Pleiades with older field dwarfs can elucidate such processes.  For example, a higher occurrence of planets around young cluster stars compared to field stars would point to subsequent photoevaporation of hydrogen envelopes and decreasing planet radii that renders older planets more difficult to detect.  A lower occurrence of planets could be explained by subsequent inward migration of planets to more detectable orbits.

Because of Malmquist (limiting magnitude) bias and the Galactic latitude ($b \sim 13^{\circ}$) of the field of the \kep{} ``prime'' mission, its target stars are primarily middle-aged dwarfs and evolved (subgiant) stars \citep{Gaidos2013}.  The successor K2 mission is observing fields near the ecliptic place which include several young clusters, OB associations, and star-forming regions, i.e. the Hyades, Pleiades, Praesepe, $\rho$ Ophiucus, Upper Scorpius, and Taurus clusters.  The Zodiacal Exoplanets in Time (ZEIT) project is confirming and characterizing candidate transiting planets detected by K2 around young cluster stars.  We previously reported a Neptune-size planet around a $\sim$650~Myr-old Hyades star \citep{Mann2016} \citep[see also ][]{David2016}) and a "super-Neptune" around a $\sim11$ Myr-old member of the Upper Scorpius OB association \citep{Mann2016B}.  Here we report the discovery and validation of a $\approx$2.3-Earth-radius (\rearth{}) ``super-Earth''-size planet on a 8.2-day orbit around \host{}, a K-type dwarf star in the vicinity of the Pleiades.  This star was selected on the basis of multiple Guest Observer target proposals for Campaign 4, which includes both the Pleiades and a portion of the Hyades cluster.  In Section \ref{sec.observations} we describe the K2 photometry of \host{} and our follow-up spectroscopy and imaging.  In Section \ref{sec.star} we describe our analyis and derivation of the host star parameters; in Section \ref{sec.planet} that of the planet parameters.  In Section \ref{sec.discussion} we discuss the ambiguous nature of \host{} and the implications of our planet detections --- or lack thereof --- for the planet population in the Pleiades cluster.

\section{Observations and Data Reduction} 
\label{sec.observations}

\indent 
\subsection{Kepler Photometry} 

Lightcurves processed by the \citet{Vanderburg2014} pipeline were downloaded from the STScI MAST database and subjected to secondary treatment.  Lightcurves were normalized and a Lomb-Scargle periodogram power spectrum \citep{Scargle1982} was generated.  To remove periodic variability due to spots and rotation and enhance the detection of transits, significant peaks were identified and any signal with a period within 10\% of these peaks was filtered from a fast Fourier transform.  The lightcurve was then further filtered by a median filter with a sliding window of 1 day, a robust standard deviation was calculated using Tukey's biweight function \citep{Tukey1977} and anomalous positive excursions $>3\sigma$ were removed.  A linear regression of the flux vs. motion along the ``arc'' of the K2 pointing error \citep{Vanderburg2014} was re-performed and subtracted.  

\begin{figure}
 \includegraphics[width=\columnwidth]{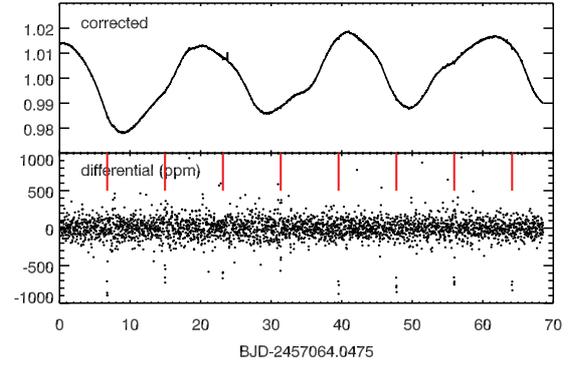}
 \caption{Top: Vanderburg pipeline-corrected and normalized lightcurve of \host, showing the $\sim$3\% peak-to-peak rotational variability.  Bottom: differential lightcurve constructed by "robustly" fitting a quadratic to the corrected data within moving a 0.5-day window centered on each data point and subtracting the predicted value at that point.  Red fiducials mark the transits of the planet described here.}
 \label{fig.lc}
\end{figure}

To identify potential transiting planet signals, an initial box-least-squares (BLS) search \citep{Kovacs2002} was performed.  The search was performed over a period range of 1-25 days and with a box width (transit duration) that ranged from 0.5 to 2 times the nominal transit duration for the estimated density of the host star, a zero impact parameter, and a circular orbit.  Significant (false alarm probability $< 0.01$) peaks in the initial BLS power spectrum were identified and an individual BLS search with a finer mesh of period and duration values was performed around each peak.  An F-test (ratio of $\chi^2$ values) was performed to evaluate whether the transit model was a significantly better fit to the data than the no-transit (flat) model.  Finally, the the mean ratio of odd- to even-numbered transits, an important indicator of false positives produced by eclipsing binaries, was calculated.  Figure \ref{fig.lc} shows the \citet{Vanderburg2014}-corrected lightcurve and a differential version exhibiting the individual transits. 

We ran an additional BLS search with a different handling of stellar variability. We first applied a median filter with a window of 2.5 hours. We removed anomalous points as before, then fed the filtered and flatted light curve into the BLS search. This short median window smooths out the light curve but also erases or weakens shallow and/or long duration transits. If no significant peak is identified in the BLS we widened the window in increments of 0.25 hours up to one day.  If a significant signal is identified in the BLS we re-fit the stellar variability, interpolating over the transits, and fit the resulting transit using Levenberg-Markwardt least-squares minimization \citep{Markwardt2009}. Because this method produces a larger number of false positives we required the transit signal to be consistent with a non-grazing transit (with a flat bottom), a planetary size, have a duration and period inconsistent with the stellar variability, and be symmetric about the mid-point, in addition to the odd-even test performed above. 

\subsection{Low-Resolution Spectroscopy}\label{sec.lowres}

A moderate resolution ($R\approx1000$) visible-wavelength (3200-9700\AA) spectrum was obtained on UT 17 January 2016 using the SuperNova Integral Field Spectrograph \citep[SNIFS][]{Aldering2002,Lantz2004} on the UH~2.2m telescope on Mauna Kea.  The integration time was 110 sec and the observation airmass was 1.01.  Details of the spectrograph and data reduction are given in \citet{Gaidos2014} and \citet{Mann2015A}.  

\begin{figure}
 \includegraphics[width=\columnwidth]{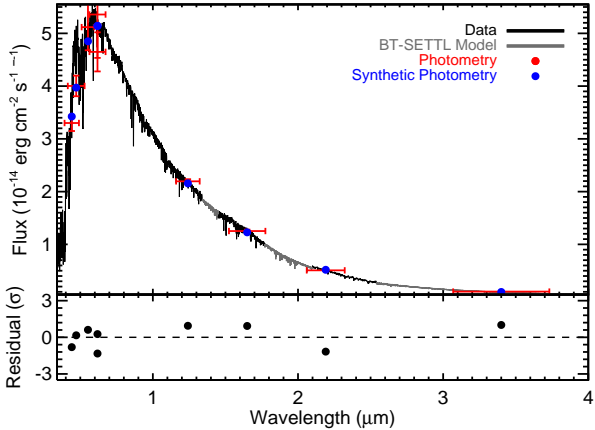}
 \caption{Flux-calibrated spectrum of \host{}, created by combining observed visible (SNIFS) and near-infrared (SpeX) spectra, filling unobservable gaps with the best-fit model spectrum, and constructing synthetic photometry from the composite spectrum to compare to the observations.}
 \label{fig.sed}
\end{figure}

A near-infrared ($JHK$, 0.8-2.4\micron) spectrum was obtained with the SpeX spectrograh on the NASA IRTF on Mauna Kea \citep{Rayner2003}.  Extraction and calibration of this spectrum was performed using the SpeXTool package \citep{Cushing2004} and corrected for telluric absorption using the spectrum of an A0 star as described in \citet{Vacca2003}.  Both the SNIFS and SpeX spectra were combined and flux-calibrated (Figure~\ref{fig.sed}) using the available photometry and the methods described in \citet{Mann2015A} and filter profiles and zero-points from \citet{Mann2015B}.

\subsection{High-resolution Spectroscopy} 

High-resolution ($R \approx 65,000$) visible-wavelength spectra were obtained on UT 26 and 29 January 2016 with the High Dispersion Spectrograph (HDS) on the Subaru telescope on Mauna Kea.  The airmass of the observations was 1.09 and 1.02, respectively.  Two different setups were used: the standard "Ub" setup covering 2980-3700\AA{} and 3820-4580\AA{} on two different CCDs, and the standard "Ra" setup covering 5110-6310\AA{} and 6580-7790\AA{}.  A 0.8" slit giving a resolution $R = 45,000$ was used for both modes.  Total integration times were 3600s and 900s with the Ub and Ra settings, respectively.  Flat fields were obtained using a quartz lamp and wavelength calibration was provided by spectra of a Th-Ar arc lamp.  Reduction (dark removal, flat-fielding, order extraction, wavelength calibration, and blaze correction) used IRAF routines developed at NAOJ and analysis was performed with custom GDL scripts.  

High-resolution NIR spectra were taken at three epochs (4 January, 3 and 4 February 2016) with the Immersion Grating Infrared Spectrometer \citep[IGRINS,][]{Park2014} attached to the 2.7m Harlan J. Smith Telescope at Mcdonald Observatory. IGRINS is a fixed-format, high-resolution ($R\simeq45,000$) NIR spectrograph with simultaneous coverage from 1.5 to 2.5\um. An A0V telluric standard was taken before or after the target each epoch. Th-Ar and U-Ne as well as dark and flat field calibration data were taken at the start of each night.  IGRINS spectra were reduced with the IGRINS pipeline package \citep[https://github.com/igrins/plp, ][]{IGRINS_plp}. This included bias, flat, and dark field corrections, and extraction of the one-dimensional spectra of both the A0V standard and target. We used the A0V spectra to correct telluric lines following the method outlined in \citet{Vacca2003}.  Pre-telluric corrected spectra of the target were retained and used to improve the wavelength solution and provide a zero-point for the RVs. Radial velocities for each IGRINS epoch were derived following the procedure outlined in \citet{Mann2016}.

\subsection{High-resolution Imaging} 

We obtained adaptive optics imaging through the $K'$ filter ($\lambda_c = 2.124$~\micron{}, $\Delta \lambda = 0.31$~\micron{}) with the Near-Infrared Camera (NIRC2) at the Keck-2 telescope on Mauna Kea during the nights of UT 18 January and 18 March 2016.  The narrow camera (pixel scale of 0.01") was used for both sets of observations.  In addition to normal AO imaging, data for non-redundant aperture masking interferometry \citep{Tuthill2000} were obtained with the 9-hole mask on the second night.  Images were processed using a custom Python pipline:  images were linearized \citep{Metchev2009}, dark-subtracted, flattened, sky-subtracted, and co-added.  A cutout $\sim$1.5$''$ across, centered on the star, was made and inserted back into the processed image as a simulated companion. The routine generated a contrast curve by decreasing the brightness and angular separation of the simulated companion with respect to the primary, until the limits of detection ($3.5\sigma$) are reached.

\begin{figure}
 \includegraphics[width=\columnwidth]{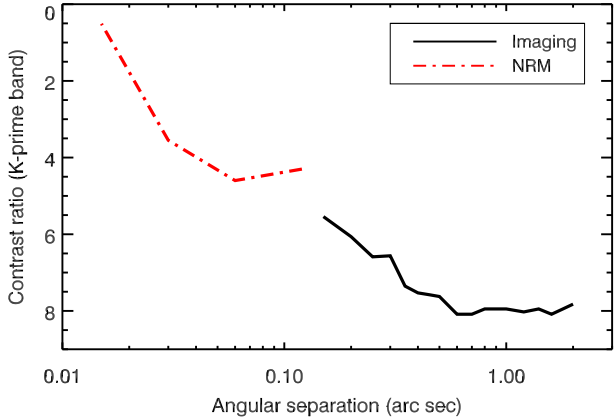}
 \caption{Detection limits (99.9\%) for sources close to \host{} placed by AO observations with Keck-2, reported as a magnitude contrast in the $K'$ passband of the NIRC2 detector.  Limits based on non-redundant aperture mapping are plotted in red.}
 \label{fig.contrast}
\end{figure}

The data obtained for non-redundant aperture masking are  essentially a collection of interferograms from pairs of apertures created in the pupil plane of the Keck-2 telescope.  Details of the data reduction are given in the appendix of \citet{Kraus2008}. Non-common path errors introduced by the atmospheric and optical aberrations were removed using the complex triple product, or closure-phase.  In the case of \host{}, the observations were paired with those of nearby EPIC~210894022.  Binary system profiles were fit to the closure phases to calculate contrast limits.  No additional sources were detected and the combined AO imaging+NRM detection limit contrast curve is plotted in Fig. \ref{fig.contrast}.  

\section{Analysis: The Host Star}
\label{sec.star}

\subsection{Pleiades Candidate Selection}  

We identified members of the Pleiades cluster using kinematic and photometric information for the candidate stars. Proper motions were gathered from the UCAC4 \citep{Zacharias2013} and PPMXL \citep{Roser2010} catalogs for all objects in the Pleiades field, and combined with our own proper motion catalog generated from the USNO-B \citep{Monet2003}, SDSS, and 2MASS \citep{Skrutskie2006} catalogs as described in \citet{Kraus2014}. We also calculated photometric distances for both a field main sequence and a Pleiades age ($\sim$100\,Myr) Padova isochrone \citep{Bressan2012} using APASS $BV$ \citep{Henden2012} and 2MASS $JK$ photometry where available, for each candidate member. We then calculated probabilities of membership in the Pleiades according to the Bayesian method described in \citet{Rizzuto2011} and \citet{Rizzuto2015} and the Pleiades kinematics of \citet{vanLeeuwen2009}. With the radial velocity from our IGRINS spectrum, \host{} was assigned a 99.6\% probability of membership in the Pleiades as opposed to a field star.    

\subsection{Kinematics and Distance} 

\begin{figure}
 \includegraphics[width=\columnwidth]{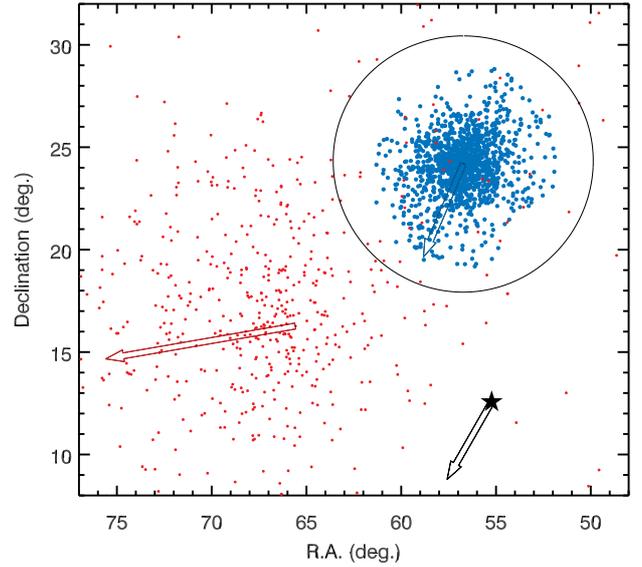}
 \caption{Location of \host{} (star) with respect to the Pleiades (blue) and Hyades (red) clusters.  Arrows show the relative magnitude and direction of proper motions.  Pleiades stars are from the catalog of \citet{Stauffer2007} and Hyades stars are from the catalog of \citet{Roser2011}.  The 13.1-pc tidal radius of the Pleiades cluster estimated by \citet{Adams2001} is plotted as a circle.}
 \label{fig.map}
\end{figure}

\host{} lies at a projected distance of 11.7$^{\circ}$. ($\approx 28$~pc) from the cluster centroid (Fig. \ref{fig.map}) and well outside the tidal radius of 13~pc \citep{Adams2001}.  No parallax is available for \host{} thus we estimated a distance assuming it is a Pleiades member and finding the distance which maximizes the probability that its space motions would be that of a cluster member.  We adopted the {\it Hipparcos}-based Pleiades $UVW$ values and uncertainties of \citet{vanLeeuwen2009}, and a Gaussian isotropic velocity dispersion with $\sigma = 0.6$~km~sec$^{-1}$ \citep{Geffert1995,Li1999,Makarov2001}.  We repeated the calculations, altering the input parameters according to normal distribution with the published standard errors and summing the distributions.  The resulting distance posterior has a mean value of 125~pc with a standard deviation of 5~pc.  This is (marginally) smaller than the VLBI distance to the Pleiades \citep[$136.2 \pm 1.2$~pc, ][]{Melis2014}. The most-probable space motions are (U,V,W) = (-5.5,-27.5,-9.2) km~sec$^{-1}$ and the expectation of the velocity offset from the nominal cluster motion is 2~km~sec$^{-1}$.

\subsection{Stellar Rotation}  
\label{sec.rotation}
The K2 lightcurve of \host{} shows pronounced ($\sim$3\% peak-to-peak) variability with a period of $\approx$20~d that we attribute to rotation and star spots.  We performed a systematic determination of the rotation periods of \host{} as well as our entire sample of Pleiades candidate members, following the procedure outlined by \citet{McQuillan2013}.  The lightcurves generated by the \citet{Vanderburg2014} pipeline were downloaded from the STScI archive, normalized by the median value, and detrended with a robust fit of a quadratic with time.  Each detrended lightcurve was then interpolated onto a regular grid set by the \kep{} long cadence of 29.4~min.  An autocorrelation function (ACF) was computed, and median-smoothed with an interval of 5.5~hr (11 points).  Maxima and minima were identified and the second and third maxima were inspected (the first maximum is at zero lag).  Because rotational variability sometimes contains a higher harmonic, the third, rather than the second maximum in the ACF may be the actual period - this is usually the larger of the two extrema.  A Gaussian function was fit to the ACF in the vicinity of this maximum to obtain a refined estimate for the period.  

\begin{figure}
 \includegraphics[width=\columnwidth]{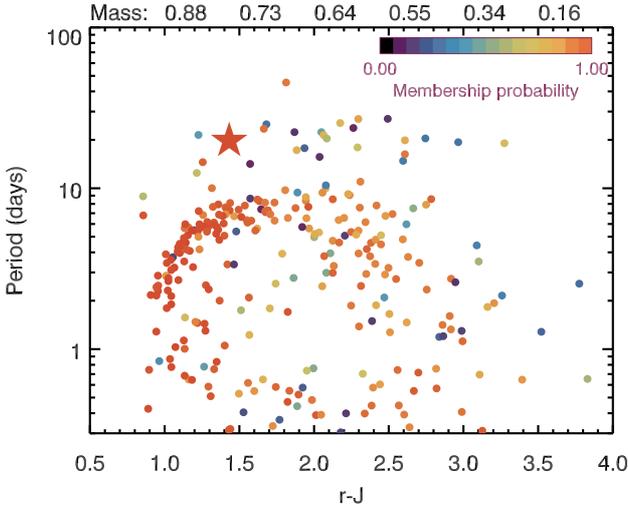}
 \caption{Rotation periods of our 384 Pleiades candidates, including \host{} (star), vs. $r$-$J$ color.  Rotation periods were determined using the autocorrelation function following the procedure of \citep{McQuillan2013}.  Membership probability is indicated by the colors of the points (see colorbar).  Equivalent stellar masses (solar units) estimated from a 120-Myr solar-metallicity isochrone from the Dartmouth Stellar Evolution Program \citep{Dotter2008} are reported for some $r$-$J$ colors.}  
 \label{fig.rotation}
 \end{figure}

The ACF-based rotational period of \host{} was found to be 19.8~d.  The highest peak in the Lomb-Scargle periodogram is at 20.0~d.  We successfully determined rotational periods for 382 other Pleiades candidates following the same method. The rotational period and $r-J$ color of these Pleiades candidates are plotted in Fig. \ref{fig.rotation}, with \host{} indicated as the star, and points color-coded according to membership probability.  Equivalent stellar masses on a 120-Myr solar-metallicity isochrone, calculated using the Dartmouth Stellar Evolution Program \citep{Dotter2008}, are also shown on the top axis.  Pleiades candidates exhibit the established bifurcation between the "slow" (few days) and "fast" ($<$1~d) rotators among K dwarfs \citep{Queloz1998,Hartman2010}, as well as a more continuous dispersion among late K dwarfs and M dwarfs.  \host{} occupies a zone occupied by more slowly rotating (20-30 day) stars with a wider range of membership probabilities.

The lightcurve of \host{} (Fig. \ref{fig.lc}) and its changing shape (i.e. the rising secondary peak) is reminiscent of that of KIC~1869793, examined by \citet{Reinhold2013}, which has a similar $B-V$ color but a longer rotation period (26.2~d).  \citet{Reinhold2013} interpret this as a manifestation of migrating spots and differential rotation.  

\subsection{Stellar Activity and Lithium}

\begin{figure} 
 \includegraphics[width=\columnwidth]{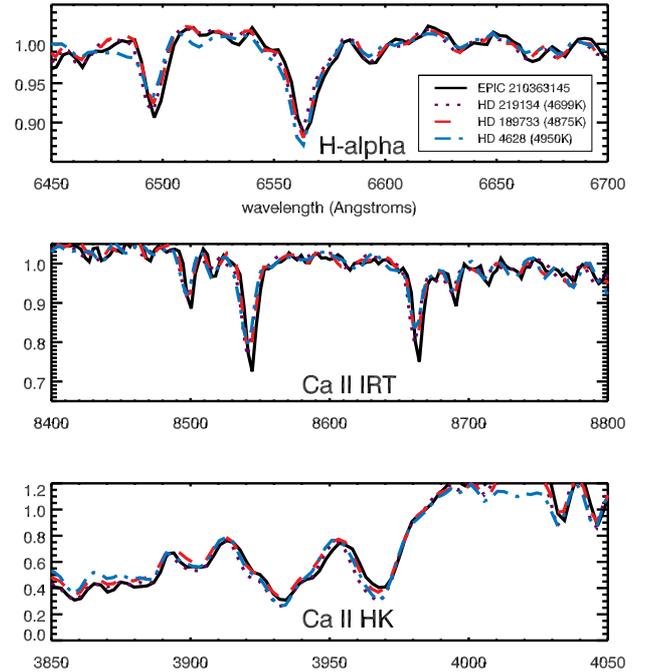}
 \caption{SNIFS spectrum of \host{} in the vicinity of the (top) 6563\AA{} Balmer $\alpha$ line of H I, (middle) the infrared triplet of Ca II, and (bottom) the H and K lines of Ca II, emission in any of  which is an indication of magnetic activity on dwarf stars.  In each region the spectrum is compared to that of three K dwarf stars with similar effective temperatures obtained by combining interferometrically measured angular radii and bolometric fluxes.  Spectra are normalized by median values and wavelengths are presented in the observed frame, in air; shifts are not removed to allow spectra to be distinguished.  Note the difference in scales between plots.}  
 \label{fig.ha}
\end{figure}

Neither the SNIFS or HDS spectrum of \host{} show evidence of H$\alpha$ in emission; instead, as is characteristic of K dwarfs it is in absorption (Fig. \ref{fig.ha}), with an equivalent width (EW) of $-1.08 \pm 0.31$\AA{} measured from the SNIFS spectrum.  We compared the spectrum of \host{} in the vicinity of the H$\alpha$, Ca II infrared triplet, and Ca II H and K lines with those of three inactive K dwarfs with similar effective temperatures accurately established by combining interferometrically measured angular radii and bolometric fluxes \citep{Boyajian2012,Boyajian2015}.  The comparison (Fig. \ref{fig.ha}) shows that the spectra are almost indistinguishable and that of \host{} shows no filling of the cores of these lines as expected for an active, Pleiades-age star.  The H$\alpha$ EW of \host{} is indistinguishable from the three stars (-0.81, -0.81, -0.89 and -0.89 $\pm 0.01$\AA{}), and this is not sensitive to the exact \teff{}.

\begin{figure} 
 \includegraphics[width=\columnwidth]{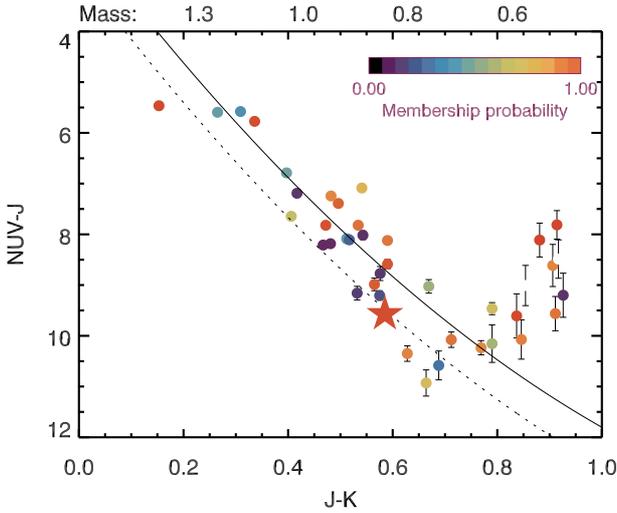}
 \caption{Near ultraviolet NUV-$J$ vs. $J$-$K$ colors of Pleiades candidates, with \host{} plotted as the star.  Equivalent stellar masses (solar units) based on a 120-Myr solar-metallicity DSEP isochrone are given for some colors.  The solid curve is the 120 Myr-locus predicted by the \citet{Findeisen2011} relation based on observations of nearby co-moving groups, assuming zero extinction.  The dashed curve is the same but assumes $A_V = 0.12$ and the extinction coefficients of \citet{Yuan2013}.}  
 \label{fig.galex}
\end{figure}

\host{} has a counterpart near-ultraviolet (NUV, $\sim2300$\AA) source in the GALEX DR-5 catalog \citep{Bianchi2011} with an AB magnitude of $19.96 \pm 0.10$.  Adopting extinction coefficients $R(\rm NUV) = 7.24$, $R(V) = 3.2$, and $R(J) = 0.72$ \citep{Yuan2013} and an extinction of $A_V = 0.12$ \citep{Guthrie1987}, the intrinsic NUV color should be $m_{\rm NUV}-J = 8.79$.  This places it directly on the locus of AB Doradus stars in the NUV$-J$ vs. $J-K_s$ color-color diagram constructed by \citet{Findeisen2011}.  The age of the AB Dor moving group is probably  similar to that of the Pleiades \citep{Luhman2005,McCarthy2014}.  We identified NUV source counterparts for 40 other Pleiades candidates from the GALEX All-sky Imaging Survey (AIS) and these are plotted vs $J-K$ in Fig. \ref{fig.galex}. Many Pleiades stars with $J-K < 0.8$ fall along a locus that is consistent with the prediction of \citet{Findeisen2011} for an age of 120~Myr based on observations of nearby co-moving groups.  \host{} and some other stars fall below that curve, but can be reconciled assuming Pleiades-like extinction.  In contrast, detected M dwarfs ($J-K>0.8$) are more UV-luminous than the \citet{Findeisen2011} prediction.  The locus is {\it not} an artifact of the detection limit  of the AIS.  For a typical limiting $m_{\rm NUV} \approx 21$ \citep{Bianchi2014}, the limiting NUV$-J$ for the Pleiades is $\ge 15$.   

There is no significant absorption at the 6708\AA{} resonant doubled of Li I in the spectrum of \host.  We placed an upper limit of 5 m\AA{} on the EW.  Using the curves of growth calculated by \citep{Lind2009}, this EW corresponds approximately to a limit on lithium abundance (with respect to the solar value) of $\log A({\rm Li}) < 0.4-0.5$ (\teff{} = 5000K, \logg{} = 4.5) or $< -0.30--0.1$ (\teff{} = 4500K, \logg{} = 4.5), depending on whether the assumption of local thermodynamic equilibrium holds.  This is below the nominal distribution for Pleiades stars \citep[e.g.,][]{Somers2015}.   

\subsection{Stellar Parameters} 
\label{sec.parameters}

We first estimated \teff{} based on $B-V$ and $V-K$ colors with $BV$ magnitudes from Data Release 9 of the APASS survey and $K$ magnitudes from 2MASS, and the \teff{}-color relations of \citep{Boyajian2012}.   We assumed a Pleiades metallicity [Fe/H] = $0.02 \pm 0.03$ \citep{Soderblom2009,Funayama2009}.  Assuming a typical Pleiades extinction of $A_V = 0.12$ and using reddening coefficients from \citet{Yuan2013}, we estimated \teff{} of $4670\pm250$ and $4910\pm100$ from $B-V$ and $V-K$, respectively.  The convolved probability distribution has a mean of $4870\pm90$K.  If extinction is negligible the mean is $4790\pm90$K.  If the star is actually metal-rich (see below), the derived \teff is hotter, but only by 20-30K.   

We made an independent estimate of \teff{} based on comparing our combined visible-wavelength (SNIFS) and near-infrared (SpeX) spectra with stellar models (Section~\ref{sec.lowres}). The combined spectrum was compared to the G\"{o}ttingen spectral library constructed using PHOENIX in spherical mode \citep{Husser2013}.  The spectrum was re-sampled at higher resolution and uniformly with logarithmic wvaelength.  Medium-resolution (R=10,000) model spectra were convolved with a Gaussian to the approximate resolution of the observed spectrum and re-sampled at the same wavelengths.  Model wavelengths were converted from vacuum to air using the formula of \citet{Morton1991} and the maximum of the cross-correlation function with the observed spectrum used as a pseudo-Doppler shift.  The shifted spectrum was then used to compute $\chi^2$ for each possible model.  Multiple regions of the spectrum that were previously identified as problematic or contaminated by telluric lines were excluded \citep{Gaidos2014}.  The 25 best-fit models were then used to construct 10,000 linear interpolations to compare with the model via $\chi^2$.  

The best-fit interpolated model with no correction for reddening has \teff{}=4770~K, \logg{}=3.72, and [M/H] = 0.5 $(\chi^2_{\nu} = 1.24$).  When dereddened ($E_{B-V} = 0.04$) was applied the best-fit model has \teff{}=4870~K, \logg{}=3.4, and [M/H] = -0.5.  The temperatures are within one standard deviation of the values derived from photometry.  We calculated 95\% confidence intervals for these values based on the distribution of $\chi^2$ values, and found a range of $\pm 90$K for \teff{} but essentially no constraint for \logg{} or metallicity, a consequence of the correlation between these two parameters in determining the strengths of lines in th spectra of K dwarfs.

The high-resolution spectrum obtained with HDS on the Subaru telescope was compared with the version 2.0 G\"{o}ttingen library of PHOENIX model spectra \citep{Husser2013}.  The resolution of the grid was 100K in \teff{}, 0.5 dex in [Fe/H], and 0.5 in \logg{}.  The comparison was order-by-order.   Spectra were interpolated onto a logarithmic wavelength grid to facilitate Doppler shifting, and normalized and de-trending by a cubic spline fit.  Model spectra wavelengths were converted to values in air using the relation of \citet{Morton1991}, the spectra were convolved with a series of Gaussians of different widths to account for rotational broadening, cross-correlated and shifted with respect to the observed spectra, and the results compared to the observations by a $\chi^2$ calculation, after removal of any second-order trend, with Doppler shift and broadening as the two free parameters.  The best-fit model has \teff{} = 4900K, [Fe/H] = 0, and \logg{} = 5.0, consistent with our results using photometry and low-resolution spectra.  The best-fit broadening is 6.6~\kms{}, equal to the instrument resolution.  Thus any rotational broadening is less than a few \kms{}, further evidence that the star is a slow rotator (Sec. \ref{sec.rotation}).  An independent, line-by-line (Fe I and Fe II) analysis of the HDS spectrum was performed using the methods described in \citet{Takeda2002} and \citet{Hirano2014}.  The derived parameters are similar: \teff{} = $4970 \pm 45$K, [Fe/H] = $0.29\pm 0.06$, \logg{} = $4.47 \pm 0.13$, and $v \sin i = 4.0$~\kms{} (for a microturbulence parameter of $\xi = 1.5$~\kms).  We adopt these values for the stellar parameters in Table \ref{tab.system}.  

The bolometric flux was determined by integrating over the flux-calibrated spectrum: $f_{\rm bol} = 5.06 \pm 0.16 \times 10^{-13}$ W~m$^{-2}$.  Adopting the spectroscopic value for \teff{} (4970K) and the kinematic distance $125 \pm 5$~pc, we estimated the radius to be $R_* = 0.67 \pm 0.03$\rsun.  Employing the empirical relation with \teff{} constructed by \citet{Boyajian2012} and considering all errors, including the intrinsic scatter in the relation itself, we found $R_* = 0.76\pm0.03$\rsun{}.  The (photometric) distance at which the two radius estimates are in agreement is  $141\pm6$~pc, identical to within error of the VLBI-based Pleiades distance.  The stellar luminosity is then $M_K = 4.05\pm0.10$ and its mass, based on the empirical relation developed by \citet{Henry1993}, is $0.78 \pm 0.12$\msun.  The bolometric luminosity, uncorrected for any extinction, is $0.316\pm0.026L_{\odot}$.  The resulting stellar \logg{} is $4.58\pm0.07$, within $1\sigma$ of the spectroscopic value.

\section{Analysis: The Planet}
\label{sec.planet}

\subsection{Light Curve Fitting}

We fit the K2 light curve with a Monte Carlo Markov Chain (MCMC) as described in \citet{Mann2016}, which we briefly summarize here. We used the \texttt{emcee} Python module \citep{Foreman-Mackey2013} to fit the model lightcurves produced by the \textit{batman} package \citep{Kreidberg2015} using the \citet{MandelAgol2002} algorithm. Following \citet{Kipping2010} we over-sampled and binned the model to match the 30\,minute cadence. We sampled the planet-to-star radius ratio ($R_P/R_*$), impact parameter ($b$), orbital period ($P$), epoch of the first transit mid-point ($T_0$), mean stellar density $\rho_*$, and two limb-darkening parameters ($q_1$ and $q_2$), and two parameters that describe the orbital eccentricity ($e$) and argument of periastron ($\omega$): $\sqrt{e}\sin{\omega}$ and $\sqrt{e}\cos{\omega}$.  We assumed a quadratic limb darkening law and used the triangular method of \citet{Kipping2013} to sample the 2-D parameter space. We applied a prior on limb darkening derived from the \citet{Husser2013} atmospheric models, calculated using the LDTK toolkit \citep{Parvianen2015}. 

We ran two separate MCMC analyses, the first ("Fit 1") with $\sqrt{e}\sin{\omega}$ and $\sqrt{e}\cos{\omega}$ fixed at zero and $\rho_*$ under a uniform prior, and the second ("Fit 2") with both $\sqrt{e}\sin{\omega}$ and $\sqrt{e}\cos{\omega}$ allowed to explore [0,1] under uniform priors, but with a Gaussian prior on $\rho_*$ derived from our stellar parameters in Section~\ref{sec.parameters}.  All other parameters were explored with only physical limitations (e.g., $P>0$) and uniform priors. MCMC chains were run using 200 walkers, each with 100,000 steps after a burn-in phase of 10,000 steps. 

\begin{table}
\begin{center}\label{tab:planet}
\caption{Transit fit parameters} 
\begin{tabular}{l l l l l l l l l l l }
Parameter & Fit 1$^a$ & Fit 2$^a$  \\
\hline
Period (days) & $8.199833^{+0.000681}_{-0.000691}$  & $8.199796^{+0.000683}_{-0.000719}$ \\
$R_P/R_*$ & $0.0273^{+0.0016}_{-0.0008}$  & $0.0278^{+0.0022}_{-0.0011}$ \\
T$_0$ (BJD-2400000) & $57070.80576^{+0.00174}_{-0.00174}$ & $57070.80596^{+0.00190}_{-0.00179}$ \\
Impact Parameter & $0.32^{+0.29}_{-0.22}$ & $0.46^{+0.24}_{-0.30}$ \\
Density (Solar) & $2.5^{+0.5}_{-1.0}$ & $1.8^{+0.3}_{-0.3}$ \\
Duration (hours) & $2.62^{+0.08}_{-0.08}$ & $2.40^{+0.35}_{-0.75}$ \\
Inclination (degrees) & $89.2^{+0.5}_{-1.0}$ & $88.7^{+0.8}_{-0.8}$ \\
a/R$_*$ & $23.3^{+1.4}_{-3.6}$ & $23.2^{+6.4}_{-2.3}$ \\
Eccentricity & 0 (fixed) & $0.14^{+  0.22}_{-0.09}$ \\
$\omega$ (degrees) & 0 (fixed) & $89^{+ 44}_{-44}$ \\
\hline
\end{tabular}
\end{center}
$^a$ Fit 1 is done with $e$ and $\omega$ fixed to 0 and a uniform prior on $\rho_*$. Fit 2 is done with a Guassian prior on $\rho_*$ from Section~\ref{sec.planet} but no constraints on $\sqrt{e}\sin{\omega}$ and $\sqrt{e}\cos{\omega}$. \\
$^b$ BJD is reported as Barycentric Dynamical Time. 
\end{table}

\begin{figure}
	\centering
	\includegraphics[width=0.95\columnwidth]{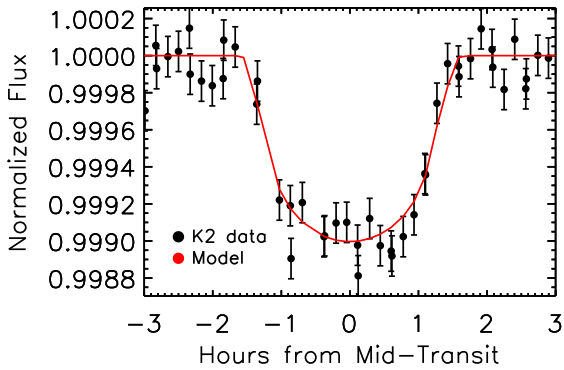} 
	\caption{Phase-folded light curve of \host\ (black) from K2. The best-fit (highest likelihood) transit model is shown in red. }
	\label{fig:transitfit}
\end{figure}

\begin{figure*}
	\centering
	\includegraphics[width=0.46\textwidth]{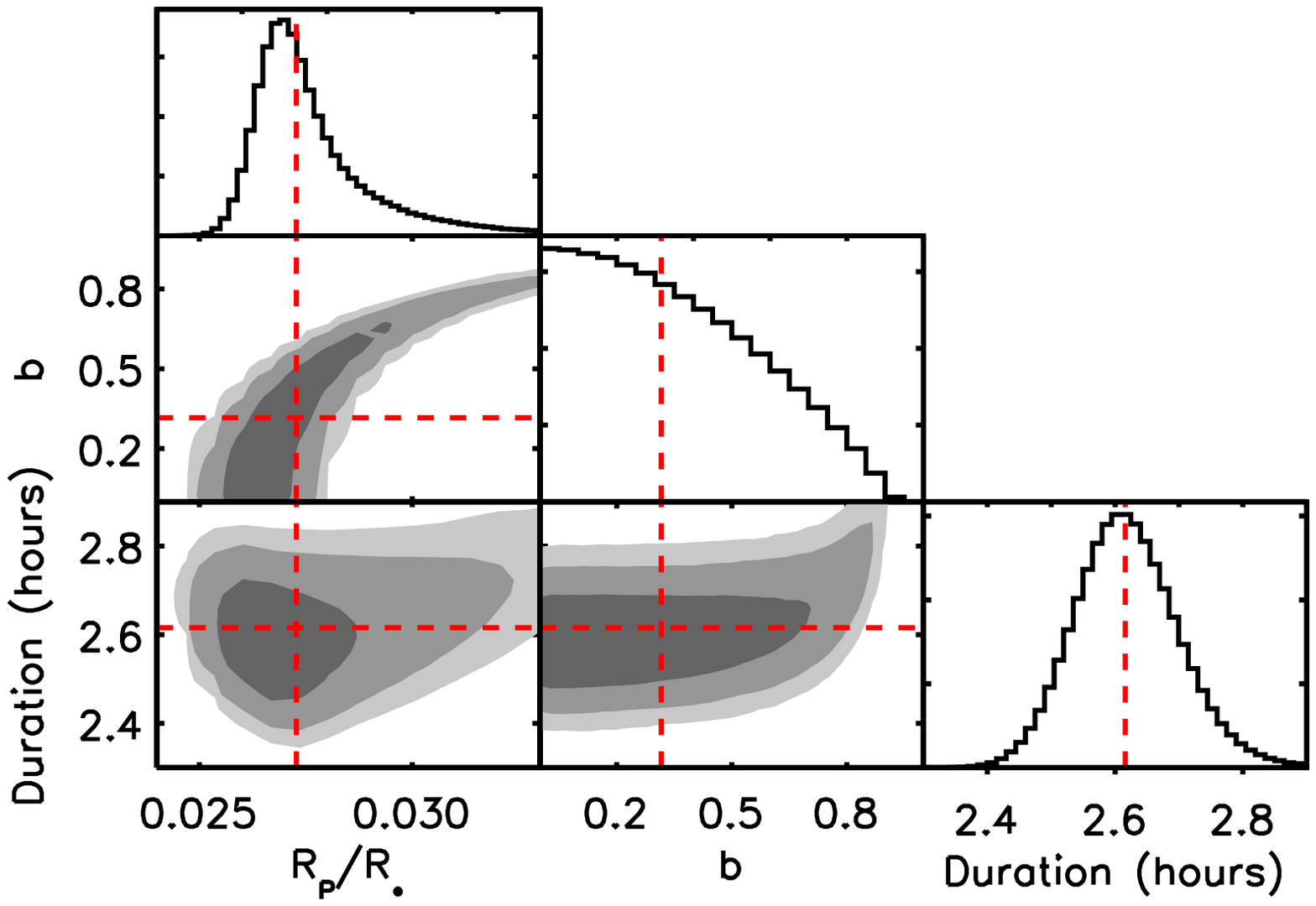} 
	\includegraphics[width=0.46\textwidth]{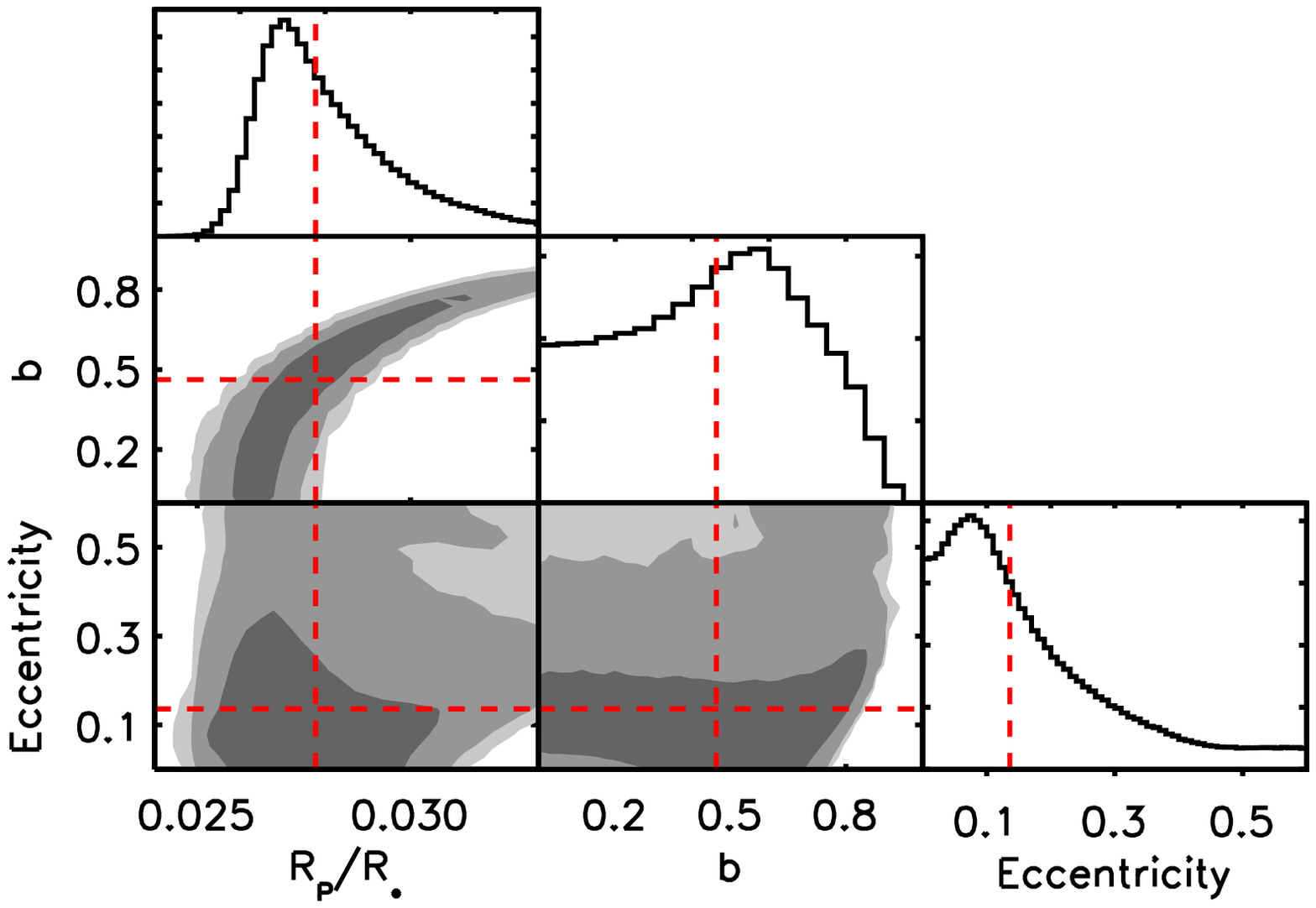} 
	\caption{Posteriors (histograms) and parameter correlations (contour plots) from our MCMC fits to the K2 light curves. Median values for each parameter are marked with red dashed lines. Grey shading covers 67\%, 95\% and 99\%, from dark to light, of the MCMC posterior. The left plot shows the posteriors when $e$ is set to zero (Fit 1), while the right is from the fit where these are allowed to float but with a density prior (Fit 2).}
	\label{fig:transitparams}
\end{figure*}

We report the transit fit parameters in Table~\ref{tab:planet}. For each parameter we report the median value with the errors as the 84.1 and 15.9 percentile values (corresponding to 1$\sigma$ for Gaussian distributions). The model light curve with the best-fit parameters is shown in Figure~\ref{fig:transitfit}. We also show posteriors and correlations for a subset of parameters in Figure~\ref{fig:transitparams}.

\begin{figure}
	\centering
	\includegraphics[width=0.95\columnwidth]{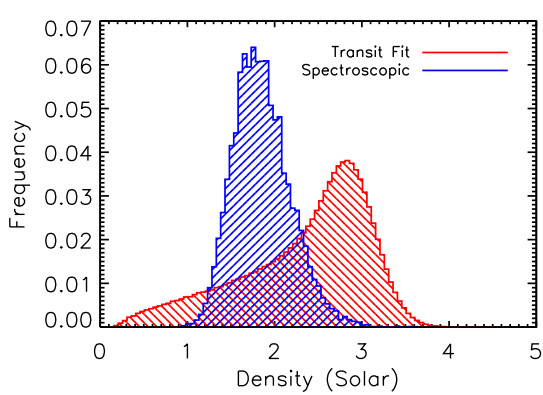} 
	\caption{Stellar density posterior from the transit fit (red) and from our stellar parameters derived independently in Section~\ref{sec.parameters} (blue). The transit fit assumes $e$=0.}
	\label{fig:dencomp}
\end{figure}

The stellar density derived from Fit 1 is consistent with our derived stellar density in Section~\ref{sec.planet}, as we show in Figure~\ref{fig:dencomp}. This is consistent with our second transit fit, which suggests a small or even zero eccentricity, and yields a density consistent with the spectroscopic/distance based estimate. 

\subsection{False-Positive Probability}

We calculated the false-positive probability (FPP) that the transit signal is produced by an unresolved background eclipsing binary using the Bayesian procedure described in \citet{Gaidos2016}.  This calculation uses as priors the TRILEGAL model (v. 1.6) of the background Galactic stellar population at the location of \host{}, and imposes constraints from transit depth, the duration of the transit, and the lack of additional sources in NIRC2-AO and NRM imaging.  The resulting FPP after 100 iterations of 10,000 Monte Carlo simulations each is $7.5 \times 10^{-7}$.  

\subsection{Radial Velocities and Planet Mass}  
\label{sec.rv}
\begin{figure}
	\centering
	\includegraphics[width=0.95\columnwidth]{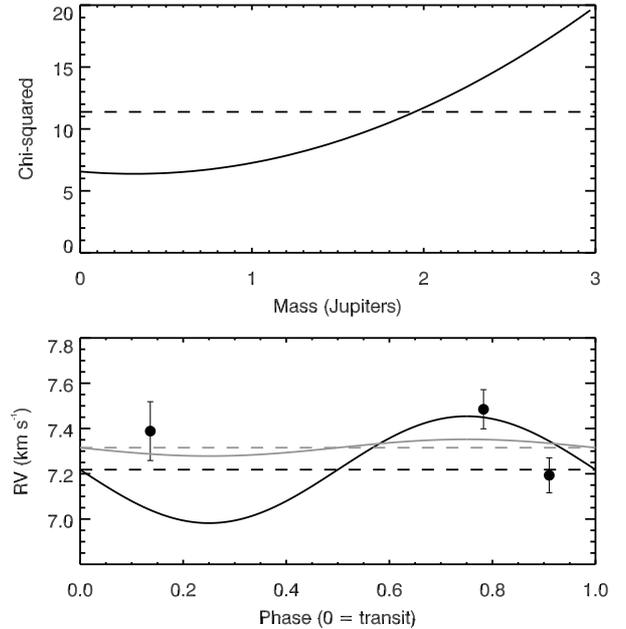} 
	\caption{Top:  $\chi^2$ of fits of a Keplerian circular, edge-on orbit with a period and phase set by the transit ephemeris to the RV data of \host{} obtained with the IGRINS spectrograph.  The single free parameter is the radial velocity of the system barycenter.  The dashed line shows the limit on $\chi^2$ corresponding to the 99\% confidence interval.  Bottom: Barycentric radial velocities phased to the orbit along with the best-fit model (solid grey) and its barycenter value (dashed grey) and 99\% upper limit model (black solid) and its barycenter value (dashed black).}
	\label{fig.rv}
\end{figure}

The three RV measurements from IGRINS are plotted vs. orbital phase (0 = transit center) and with their {\it relative} error bars in Fig. \ref{fig.rv}.  Assuming that RV variation is due exclusively to the the presence of the transiting planet, as well as a circular, edge-on orbit, we placed an upper limit on the planet mass.  For each possible mass (and corresponding Doppler RV amplitude) we fit for the only remaining free parameter -- the system barycenter velocity -- and calculated $\chi^2$.  We identified the mass for minimum $\chi^2$ (0.3 Jupiter masses $M_J$, $\chi^2 = 6.4$, $\nu = 2$) but also a much more meaningful 99\% upper limit of $1.9M_J$ based on $\chi^2$ ($\Delta \chi^2 < 5.0$).  This rules out a stellar or brown dwarf mass for the transiting object.  Although we have neglected astrophysical noise (stellar "jitter") in our analysis, this is expected to be comparatively small at infrared wavelengths \citep{Marchwinski2015} and much less than the typical measurement error of $\sim$100 \mps.

\section{Discussion}
\label{sec.discussion}

\subsection{The Nature of \host{} and its Planet}
\label{sec.nature}

Our estimates of the properties of the \host{} system are summarized in Table \ref{tab.system}.  \host{} was identified as a Pleiades member with high confidence based on its proper motion and photometric distance, but our detailed analysis of the star's properties contract this.  The star is location at approximately the same distance as the cluster, but it is $\approx$28 pc, about twice the tidal radius, from the cluster center.  Assuming the photometric distance is correct, the space motion of \host{} is actually distinct from that of the Pleiades by $5.4 \pm 0.9$ \kms{}, or about 9 times the cluster's velocity dispersion.  The star's relative motion is at a right angle ($92 \pm 12^{\circ}$) to the vector between the star and the cluster center, thus \host{} cannot be a "runaway" from the Pleiades.  Although the star is rotationally variable and a NUV source, its rotation period is 20~d, twice as long as the most slowly rotating Pleiades stars, and its spectrum lacks H$\alpha$ emission.  Finally, it is significantly more metal-rich (by $\approx$0.26~dex) than the typical Pleiades member.

\begin{table}
\caption{Properties of the EPIC~210363145 System} 
\label{tab.system}
\begin{tabular}{ll}
Parameter & Value\\ 
\hline
Observed: & \\
R.A. (J2000) & 03h 40m 54.82s\\
Dec. & 12$^{\circ}$\,34'\,21.4''\\
$\mu_{\rm RA}$ & $+23.5 \pm 1.3$ mas yr$^{-1}$\\
$\mu_{\delta}$ & $-37.9 \pm 1.7$ mas yr$^{-1}$\\
$(U,V,W)$ & ($-5.4\pm0.6$,$-28.7\pm1.0$,$-9.5\pm0.9$) \kms{} \\
$B$ & $13.229 \pm 0.030$ \\
$V$ & $12.158 \pm 0.093$ \\
$g$ & $12.717 \pm 0.019$ \\
$r$ & $11.813 \pm 0.087$ \\
$i$ & $11.531 \pm 0.060$ \\
$J$ & $10.384 \pm 0.020$ \\
$H$ & $9.910 \pm 0.023$ \\
$K_s$ & $9.799 \pm 0.018$ \\
\hline
Inferred (Star): & \\
Dist. (kinematic) & $125\pm7$~pc \\
Dist. (photometric) & $141\pm6$~pc \\
\teff{} & $4970\pm45$~K\\
Fe/H & $+0.29\pm0.06$~dex \\
\logg{} & $4.47\pm0.13$ \\
$R_*$ & $0.76\pm0.03$~\rsun \\
$M_*$ & $0.80\pm0.12$~\msun \\
$L_*$ & $0.316\pm0.026 L_{\odot}$ \\
$P_{\rm rot}$ & 19.8~days \\
$v \sin i$ & 4.0~\kms{} \\
Age & 0.4-1~Gyr? \\
\hline
Inferred (Planet): & \\
Planet radius & $2.30 \pm 0.16$~\rearth{}\\
Planet mass & $< 1.9 M_J$ \\
Orbital period & $8.1998 \pm 0.0007$~days \\
Orbital inclination & $88.7 \pm 0.8$~deg. \\
Orbital eccentricity & $0.14^{+0.22}_{-0.09}$~deg.\\
Stellar irradiance & $58 \pm 7 I_{\oplus}$ \\
\hline
\end{tabular}
\end{table}

 The star does not appear kinematically related to any other known cluster or nearby young moving group.  The AB Doradus moving group may be related to the Pleiades via a similar age \citep{Luhman2005} and kinematics \citep{Ortega2007}, but its known members are much closer than \host{}.  Nevertheless, \host{} is moving comparatively slowly ($\Delta v \approx$16~\kms{} and $UVW$ = +3.1,-15.3,-3.0 \kms{}) with respect to the Local Standard of Rest as defined by \citet{Coskonoglu2011} and is probably young.  According to the age-kinematic relations of \citet{Nordstrom2004}, such motion is characteristic of an age of $< 1$~Gyr.  Curiously, most of the motion w.r.t. the LSR in the direction of Galactic rotation ($V$) and not $U$, but this cannot be due to differential Galactic rotation alone over reasonable scales, i.e. if the Oort constants $A + B$ are $\sim 2.5$ \kms~kpc$^{-1}$ \citep{Feast1997}.
 
 The star's 20-day rotation period places it between the between the 450~Myr and 1~Gyr gyrochrones of \citet{Barnes2007}, indicating a young but not Pleiades-like age.  The high metallicity and slow rotation nicely explain the absence of lithium in \host{}.  Lithium depletion is vastly accelerated in metal-rich pre-main sequence stars because the higher opacity deepens the base of the convective zone, mixing Li to higher temperatures where it is destroyed.  The predicted effect for ZAMS stars is $\ge$1 dex decrease in Li for each 0.1 dex increase in [Fe/H] \citep{Somers2014}, and lithium would have been depleted to undetectable levels by a Pleiades-like age.  Lithium abundance also correlates with rotation \citep[e.g.,][]{Soderblom1993,Chaboyer1995}, and the relationship appears to depend on stellar mass and switch sign at $M_* \approx 1M_{\odot}$, becoming positive (and stronger) for stars with \teff{} $<5500$K.  \citet{Somers2015} proposed that in cooler stars, pre-main sequence destruction of lithium is extremely sensitive to the central temperature and hence interior structure of the star, and is inhibited by inflation of the stellar radius.  If radius inflation is related to magnetic fields \citep{Feiden2013}, which are in turn related to stellar rotation, then more rapidly rotating, magnetically active stars will have more lithium, as is observed.  Conversely, a very slow rotator such as \host{} will be magnetically inactive, hhave experienced no inflation, and have suffered the full lithium depletion predicted for such objects \citep{Somers2015}.
 
\host{} emits detectable flux at NUV wavelengths consistent with a Pleiades age but no H$\alpha$.  The reliability of the NUV detection by GALEX ($m \approx 20$) is about 99\%. \footnote{http://www.galex.caltech.edu/researcher/techdocs.html}  The de-reddened $m_{\rm NUV} - J$ color of \host{} is 8.8 and about 1.9 magnitudes "bluer" (NUV brighter) than an extrapolation of the locus of inactive field stars described by \citet{Ansdell2015}.  An M dwarf star with such a $m_{\rm NUV} - J$ color typically has H$\alpha$ emission with an equivalent width (EW) of $\sim 4$\AA{} (note that \host{} is an early K dwarf).  However, the expected H$\alpha$ emission may be smaller considering that, since \host{} does not appear to be a Pleiades member and does not lie along the line of sight to the cluster center, the extinction may be negligible.  In this case, the star's $m_{\rm NUV}-J$ is redder ($\approx 9.6$) and only 1.1 magnitudes "bluer" than the locus of inactive stars.  The expected EW of H$\alpha$ emission is $\sim2$\AA, however the scatter spans values from zero to twice this value.  This scatter could be the result of multi-year-long cycles of stellar activity.  For example, observations in 1988 by \citet{Soderblom1993} identified several bona fide Pleiades members lacking significant H$\alpha$ emission, including Trumpler HII3179 (HD~24194).  Observations by \cite{White2007} 14 years later found HD~24194 to have H$\alpha$ emission with an EW of 3.16\AA.HD~24194).  Likewise, the absence of H$\alpha$ emission from \host{} may be due to the 8+ years between the epochs of the GALEX (pre 2008) and SNIFS observations.

We ruled out two alternative explanations for the absence of H$\alpha$ emission:  GALEX confusion with an unrelated background source, or a white dwarf companion which has UV flux but no H$\alpha$ emission.  We calculated the probability of source confusion by identifying all GALEX DR5 AIS+MIS sources with $m_{\rm NUV} < 20$ within 10 arc min of the position of \host{} and multiplying this count by the ratio of the solid angle of the GALEX point response function, conservatively approximated as a circle of radius 7" \citep{Morrissey2007}, to the query area.  The result is a confusion probability of $1 \times 10^{-3}$.  Even if we include {\it all} detected sources regardless of NUV magnitude, the probability only rises to 2\%.  A white dwarf is ruled out by our RV data (Sec. \ref{sec.rv}). 

To coherently constrain the properties of \host{}, we performed a global parameter inference by comparing our observations to a grid of stellar evolution models. Model calculations were performed with an updated version of the Dartmouth stellar evolution code \citep{Dotter2008,Feiden2013} in which a number of improvements were made to more accurately model the properties of pre-main sequene stars (Feiden, in review). The grid of non-magnetic models covered a mass range of 0.1-0.9\msun{} with a resolution of 0.02\msun{} and a metallicity range of -0.5 to +0.5 dex at a resolution of 0.1 dex.  The Monte Carlo Markov Chain (MCMC) method implemented as {\tt emcee} \citep{Foreman-Mackey2013} was used to explore possible values of stellar mass, metallicity [M/H], age, and distance and find the combination that most closely reproduces the stellar \teff{}, bolometric flux $F_{\rm bol}$, and mean density $\rho_*$ using the likelihood function described in \citet{Mann2015A}.  We applied Gaussian priors on distance ($130\pm20$~pc) and metallicity ($+0.3\pm0.1$~dec) and uniform priors to mass (0.1-0.9\msun) and age (0.1~Myr-14~Gyr).  The MCMC calculation was performed with 300 walkers initially assigned to random positions in parameter space.  An initial ``burn-in'' sequence of 2000 iterations was used to seed a second sequence of 3000 iterations and then discarded.  The second sequence was used to calculate the posterior probability distributions for the stellar parameters (chains were not thinned). Convergence was evaluated by visually inspecting trace plots for all 300 chains to identify any systematic trends or noticeably poor mixing.  In addition, we verified that the median acceptance fraction among the ensemble of chains was 25-50\% and the length of each chain was at least five times the auto-correlation length.

Our model comparison found the most probable properties for \host{} to be $M_* \ge 0.89$\msun, $R_* = 0.78$\rsun, [M/H] = +0.26~dex, a distance of 143~pc, and an age of 50~Myr, although a wide range of older ages is allowed.  We do not report formal uncertainties with these values as they are highly correlated and the preferred mass is at the edge of the model grid.  The values are broadly consistent with the spectroscopically- and empirically-determine values, in part due to the priors, but the models assign the star a higher mass and slightly greater radius and distance.  At a given \teff{}, $R_*$ increases with [M/H], requiring a larger distance to reproduce $F_{\rm bol}$.   For a given $M_*$, higher [M/H] causes a {\it decrease} in \teff{}, thus higher metallicity increases the required $M_*$ to reproduce the observed \teff{}.  However, the effect of metallicity was not included in the empirical relations of \citet{Boyajian2012} and \citet{Henry1993} due to the lack of appropriate calibrator stars.  A larger stellar radius means of course that the planet is also proportionally larger.  

\begin{figure}
	\centering
	\includegraphics[width=0.95\columnwidth]{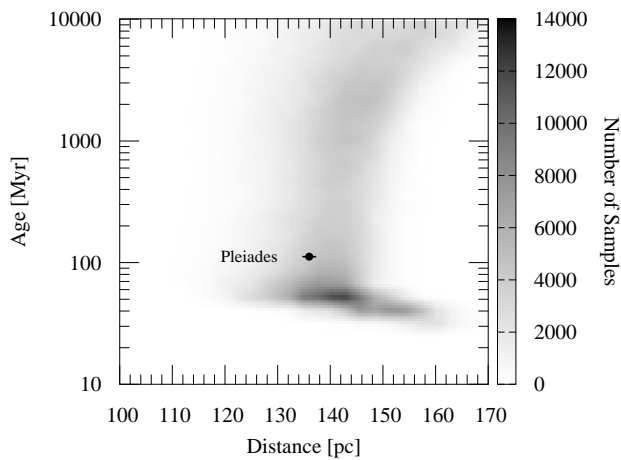} 
	\caption{Joint posterior distribution of age and distance of \host{}, calculated by MCMC inference of stellar parameters performed by comparing \teff{}, [Fe/H], and bolometric flux to the predictions of a stellar evolution model.  The location of the Pleiades ($112 \pm 5$~Myr, $136 \pm 1.2$~pc) is marked as the cross.}
	\label{fig.age_distance}
\end{figure}

Figure \ref{fig.age_distance} shows the joint posterior distribution of age and distance, marking the Pleiades mean distance and age.  Posterior distances are slightly, but not significantly farther than the center of the Pleiades and at a given distance, there is a large range of possible ages.  Most curious is the preferred age of 50~Myr, the main sequence "age zero" (ZAMS) of a star with 0.9\msun and [M/H] = +0.3.  For a given \teff{} and [Fe/H], the ZAMS is a minimum in luminosity and maximum in mean density.  Luminosity can be traded against distance when bolometric flux is fixed, and the best-fit distance is not extreme.  On the other hand, it appears the transit-based stellar density constraint, which is skewed toward higher values than the empirical/spectroscopy-based value (Fig. \ref{fig:dencomp}) drives many MCMC solutions to the ZAMS.  Nevertheless, many older ages are not excluded by our model fits.    

\begin{figure}
	\centering
	\includegraphics[width=0.95\columnwidth]{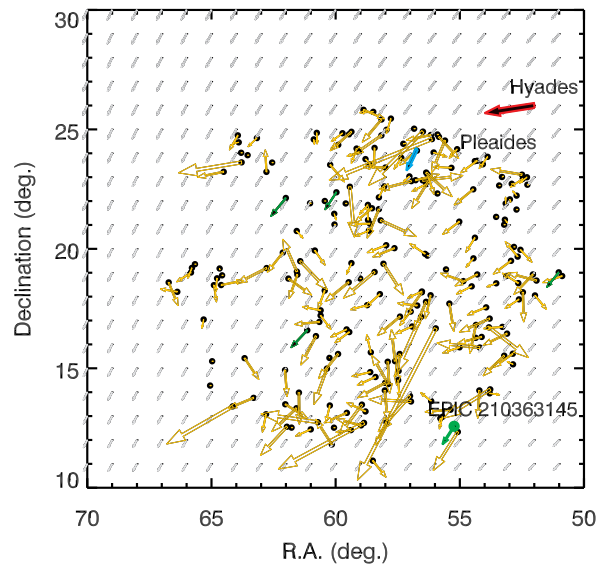} 
	\caption{Locations of 218 candidate K dwarfs observed by K2 in the C4 field which exhibit a periodic signal (e.g. from rotational variability) in their lightcurve but were not identified as members of the Pleiades or Hyades clusters.  The gold arrows indicate the direction and magnitude of proper motion, if any.  The location and PM of \host{} and the Pleiades center are shown by the green and blue arrows, respectively.  The other thin green arrows are variable K dwarf stars which have PMs within 20\% of \host.   The Hyades PM (but not location) is indicated by the red arrow.  The grid of grey arrows indicates the PM of a star in the Local Standard of Rest, i.e. the PM is due only to the space motion of the Sun.}
	\label{fig.kdwarfs}
\end{figure}

Is \host{} a loner, or instead only one of a population of variable dwarf stars with similar kinematics?  We identified 973 stars observed by K2 in the C4 field that are candidate field K dwarfs, i.e., have estimated \teff{} in \citet{Huber2016} between 4200 and 5200~K and \logg{} $>4.1$, but are not identified as members of the Hyades or Pleiades clusters by \citet{Mann2016} or this work.  We analyzed the K2 lightcurves of these stars using the ACF-based method described in Sec. \ref{sec.rotation}.  A periodic signal was identified in 218 of these stars.  Figure \ref{fig.kdwarfs} plots their locations and proper motions as well as those of \host{} and the Pleiades and Hyades clusters.  There are many rotationally variable stars with proper motions in roughly the same direction (SE) as \host{}.  The star with the largest proper motion is BD+16~502/GJ~9122A, an active K7 dwarf only 18~pc away \citep{Reid2002}.  This phenomenon is merely the result of the Sun's motion with respect to the LSR.    We calculated the this apparent motion using the LSR of \citet{Coskonoglu2011} and the photometric distance of \host{} and this is plotted as the grid of  grey arrows in Fig. \ref{fig.kdwarfs}.  To identify potential kinematic relatives of \host{}, we iteratively selected those variable K dwarfs that have PMs within $5\sigma$ of their mean, using the proper motion of \host{} as the initial value.  We identified four such stars besides \host, plotted as green arrows in Fig. \ref{fig.kdwarfs}.  All were rejected as Pleiades members, they have a wide range of rotation rates and photometric distances, and their similar proper motions appears to be coincidence.

Overall, the properties of \host{} indicate that it is unrelated to and probably older than the Pleiades, but younger ($<1$~Gyr) than most field stars.  If so, its proximity to the Pleiades in dynamical phase space is a coincidence, and a warning that membership probability calculations based on space motion and photometric distance are not immune from interlopers.  On the other hand, it is possible that this star tells us that very slow rotation is no disqualifier for a young age:  \citet{Stauffer2016} reported four F dwarf and seven M dwarf Pleiades members with anomalously low rotation rates; they attribute the phenomenon among F dwarfs to binarity, but for the M dwarfs, all of which have $P > 10$~d and some of which exhibit weak H$\alpha$ emission, they offer no explanation.   A precise parallax and proper motion from the {\it Gaia} astrometric mission will allow us to further refine the properties of this enigmatic star and its planet.   

The planet's maximum plausible mass, assuming an Earth-like rock-metal composition, is $\sim 50 M_{\oplus}$ \citep{Swift2012} and the commensurate Doppler signal is $\sim$20~\mps{}, well below our current RV limit.  However, the planet radius falls in a regime where many combinations of mass-radius measurements suggest planets have thick hydrogen-helium envelopes \citep{Rogers2015}.  In that case, adopting the nominal mass-radius relation of \citet{Weiss2014}, the planet mass is predicted to have $\sim6 M_{\oplus}$ and the Doppler signal is only 2~\mps{}.  This is possible to detect using sufficiently sensitive instruments such as ESPRESSO on the VLT \citep{Pepe2014}, provided photospheric "jitter" from spots does not overwhelm the signal.  

\subsection{Occurrence of Pleiades Planets} 

Our analysis of the K2 light curves of about 1000 candidate Pleiades members has thus far revealed only this single planet around a host star of questionable membership.  This begs the question: where are the Pleiades planets?  \kep{} discovered several thousand candidate planets around $\sim$200,000 stars, a rate of $\sim$2\%, but of course K2 observed the Pleiades for a much shorter interval of time, the photometric noise due to pointing error is larger, and Pleiades stars are more active and rotationally variable.  To address the question of whether the absence of Pleiades detections reflects a difference in planet population or is the result of lower detection efficiency, we performed numerical injection-and-recovery experiments on the K2 lightcurves assuming the null hypothesis, i.e. the same population of planets around \kep{} stars also orbits around the Pleiades.  

The transit signal from a single simulated planet on a circular orbit was injected into each of ten replicate lightcurves of the 1014 Pleiades candidates with values for \teff, radius, and mass in the EPIC catalog. Planet radii were randomly selected from the intrinsic distribution calculated by \citet{Silburt2015} for FGK dwarfs (\teff{} $>3900$K) and from \citet{Gaidos2016} for cooler M dwarfs.  Orbital periods were selected from a logarithmic uniform distribution over 3-25~d.  Transits were simulated with a random phase, an impact parameter selected from a uniform distribution and a quadratic limb-darkening law with coefficients selected from \citet{Claret2011} for the EPIC value of \teff{}.  Transits were recovered (or not) using the lightcuve processing and BLS search routine described in Sec. \ref{sec.planet}.  A planet was deemed "recovered" if the orbital period $P$ fell within $2.5\sigma$ of the true value, where $\sigma = P/N_{\rm ind}$, and $N_{\rm ind}$ is the number of independent frequencies.  ($\sigma$ is the effective spectral resolution of the BLS search).  Inspection of the distribution of deviations of recovered from ``true'' orbital periods showed that this criterion sufficed to capture essentially all detections.

The geometric probability of a circular transiting orbit was calculated and summed for all detections.  This sum represents the expected number of detections if every star has one planet with the properties in the specified ranges.  To make an actual estimate, this was multiplied by the estimated occurrence based on published \kep{} statistics.  For this step, we divided systems into FGK and M dwarfs as occurrence seems to differ markedly between spectral types \citep{Mulders2015,Gaidos2016}.  Estimates of the occurrence of planets around solar-type stars vary: \citet{Howard2012} found 0.24 per star for planets $>1$\rearth{} on 3.4-29~d orbits; \citet{Petigura2013} found 0.23, which when corrected for planets $<1$\rearth{} using the radius distribution of close-in planets is 0.28. \citet{Silburt2015} found about 0.25 for planets 1-4\rearth{}, which when corrected for smaller and larger planets is about 0.33, a value we used here.  The occurrence of planets close to M dwarfs is significantly higher:  \citet{Gaidos2016} estimated that M dwarfs host $2.2 \pm 0.3$ 1-4\rearth{} planets with $P < 180$~days, consistent with previous estimates \citep{Morton2014,Dressing2015}.  The occurrence for $P < 25$~days is 1.45.  

The predicted number of detection is 0.39 among the 411 FGK Pleiades dwarfs and 0.44 among the 603 M dwarfs, for a total of $\sim 0.83$ detections.  A separate analysis of the same set of synthetic lightcurves with a modified version of the  K2SSF pipeline \citep{Mann2016B,Vanderburg2016} predicted 1.27 detections among FGK dwarfs and 0.51 among the M dwarfs, for a total yield of 1.78.  The Poisson probabilities of having no detections are thus 0.44 and 0.17, respectively.  Thus the absence of detections among Pleiades stars is at least in part the product of the steep distribution of planet occurrence with radius and the difficulty of detecting small planets around young, photometrically variable stars, at least with our current analysis tools and the sensitivity and monitoring interval offered by K2.  Another possible factor is if the radii of Pleiades stars were systematically underestimated in the EPIC catalog, e.g. because the effect of reddening, which makes stars appear cooler and therefore smaller, was not corrected or because some Pleiades M dwarfs may not have entirely evolved to the main sequence.  The upcoming TESS mission \citet{Ricker2014}, which will survey the sky at a comparable sensitivity and with a similar time baseline, may offer only limited improvement in the situation.  Data from the PLATO mission \citep{Rauer2014} are more likely to yield advances, provided the appropriate fields are observed.

\section*{Acknowledgements}

This research was supported by NASA Origins of Solar Systems grant NNX11AC33G to EG, NSF grants AST-1518332 to R.J.D.R., J.R.G., and T.M.E., and NASA grants NNX15AD95G/NEXSS and NNX15AC89G to R.J.D.R., J.R.G., and T.M.E.  ALK acknowledges support from a NASA Keck PI Data Award administered by the NASA Exoplanet Science Institute, and from NASA XRP grant 14-XRP14 2-0106. N.N. acknowledges supports by Grant-in-Aid for Scientific Research (A) (JSPS KAKENHI Grant Number 25247026).  SNIFS on the UH 2.2-m telescope is part of the Nearby Supernova Factory project, a scientific collaboration among the Centre de Recherche Astronomique de Lyon, Institut de Physique Nuclaire de Lyon, Laboratoire de Physique Nuclaire et des Hautes Energies, Lawrence Berkeley National Laboratory, Yale University, University of Bonn, Max Planck Institute for Astrophysics, Tsinghua Center for Astrophysics, and the Centre de Physique des Particules de Marseille. Based on data from the Infrared Telescope Facility, which is operated by the University of Hawaii under Cooperative Agreement no. NNX-08AE38A with the National Aeronautics and Space Administration, Science Mission Directorate, Planetary Astronomy Program. This work used the Immersion Grating Infrared Spectrometer (IGRINS) that was developed under a collaboration between the University of Texas at Austin and the Korea Astronomy and Space Science Institute (KASI) with the financial support of the US National Science Foundation under grant AST-1229522, of the University of Texas at Austin, and of the Korean GMT Project of KASI.  Some/all of the data presented in this paper were obtained from the Mikulski Archive for Space Telescopes (MAST). STScI is operated by the Association of Universities for Research in Astronomy, Inc., under NASA contract NAS5-26555. Support for MAST for non-HST data is provided by the NASA Office of Space Science via grant NNX09AF08G and by other grants and contracts. This research has made use of the NASA/IPAC Infrared Science Archive, which is operated by the Jet Propulsion Laboratory, California Institute of Technology, under contract with the National Aeronautics and Space Administration.  This research made use of the SIMBAD and VIZIER Astronomical Databases, operated at CDS, Strasbourg, France (http://cdsweb.u-strasbg.fr/), and of NASAs Astrophysics Data System, of the Jean-Marie Mariotti Center Search service (http://www.jmmc.fr/searchcal), co-developed by FIZEAU and LAOG/IPAG.



%\bibliographystyle{mnras}
%\bibliography{references}

%%%%%%%%%%%%%%%%%%%%%%%%%%%%%%%%%%%%%%%%%%%%%%%%%%

% Don't change these lines
\bsp	% typesetting comment
\label{lastpage}
\end{document}